# Authentication of Metropolitan Quantum Key Distribution Network with Post-quantum Cryptography

YONG-HUA YANG,[1, 8] PEI-YUAN LI,[2, 8] SHI-ZHAO MA,[3] XIAO-CONG QIAN,[2] KAI-YI ZHANG,[4, 5] LIU-JUN WANG,[6] WAN-LI ZHANG,[3] FEI ZHOU,[3] SHI-BIAO TANG,[2] JIA-YONG WANG,[1] YU YU,[4, 5] QIANG ZHANG,[3, 7] AND JIAN-WEI PAN,[7]

[1]*CAS Quantum Network Co., Ltd, Shanghai 201315, China*
[2]*QuantumCTek Co., Ltd, Hefei 230088, China*
[3]*Jinan Institute of Quantum Technology, Jinan 250101, China*
[4]*Department of Computer Science and Engineering, Shanghai Jiao Tong University, Shanghai 200240, China*
[5]*Shanghai Qizhi Institute, Shanghai 200232, China*
[6]*School of Physics and Astronomy and Yunnan Key Laboratory for Quantum Information, Yunnan University, Kunming 650500, China*
[7]*Hefei National Laboratory for Physical Sciences at Microscale and Department of Modern Physics, University of Science and Technology of China, Hefei 230026, China*
[8]*These authors contributed equally to this work*

**Abstract:** Quantum key distribution (QKD) provides information theoretically secures key exchange requiring authentication of the classic data processing channel via pre-sharing of symmetric private keys. In previous studies, the lattice-based post-quantum digital signature algorithm Aigis-Sig, combined with public-key infrastructure (PKI) was used to achieve high-efficiency quantum security authentication of QKD, and its advantages in simplifying the MAN network structure and new user entry were demonstrated. This experiment further integrates the PQC algorithm into the commercial QKD system, the Jinan field metropolitan QKD network comprised of 14 user nodes and 5 optical switching nodes. The feasibility, effectiveness and stability of the post-quantum cryptography (PQC) algorithm and advantages of replacing trusted relays with optical switching brought by PQC authentication large-scale metropolitan area QKD network were verified. QKD with PQC authentication has potential in quantum-secure communications, specifically in metropolitan QKD networks.

## 1. Introduction

In recent years, experiments and practical applications of quantum key distribution (QKD) has made remarkable progress. Starting from an early proof of concept in the laboratory of more than 32 cm [1], to a later expansion to 100 km [2, 3], the maximum key distribution distance through practical optical fiber has now exceeded 500 km [4, 5]. By employing trusted relays, some QKD networks has also been tested outside the laboratory [6-14], and application demonstrations have been conducted in different fields, such as government affairs, finance, and electricity power. Further, a 4,600 km quantum secure communication network utilizing the Micius satellite and "Beijing-Shanghai backbone" intercity quantum communication network has been built [10].

QKD can provide information theoretically secure key exchange [15-17], including a quantum channel for transmitting photons and a classical channel for data post-processing. However, the unconditional security of QKD requires that the classical channel be authenticated to prevent man-in-the-middle attacks. Further, in the QKD process, basis sifting, error correction, and privacy amplification, along with the verification of the final key must be authenticated. At present, the secure authentication method used by QKD devices helps in pre-sharing symmetric keys [18]. Every paired user is required to store his/her shared authentication keys. However, the storage, synchronization, and management of a large number of key pairs increases the management and security risks of the network. Another secure authentication method is the use of post-quantum public key algorithm and public-key infrastructure (PKI) [19], where each user applies for a digital certificate from the authentication center and performs the authentication based on the post-quantum cryptography (PQC) public key algorithm. This method not only overcomes the shortage of pre-sharing symmetric keys, which arises when the number of network users is large, it can also provide quantum resistant security.

In a previous study, we adopted the lattice-based post-quantum digital signature algorithm Aigis-Sig [20], combined with PKI to achieve a high-efficiency quantum security authentication for the QKD, and thereafter, we verified the feasibility of using PQC for QKD authentication in a laboratory environment [21].

The feasibility and stability of using PQC authentication in an actual metropolitan QKD network, as well as the performance of QKD after integrating PQC, still need to be verified. Using PQC authentication, the trusted relays in the laboratory environment network can be replaced by optical switches, but its applicability to the metropolitan QKD network with more users and complex network topology also needs to be verified. In addition, in the previous experiment, the PQC algorithm was implemented using a software that ran on a PC independent of the QKD device, and the feasibility of integrating it into a QKD device needs to be verified in practice.

Based on the above unanswered questions, in this experiment, we further integrated the PQC algorithm into a commercial QKD device. In the Jinan field metropolitan QKD network, we verified the high-efficiency quantum secure authentication of QKD by using a lattice-based post-quantum digital signature algorithm Aigis-Sig combined with PKI.

## 2. Experiment and Results

The QKD system used in this experiment is a commercial device that adopts polarization encoding based on the decoy-state BB84 protocol [22]. Its repetition rate was 40 MHz, and an InGaAs gated single-photon detector with a detection efficiency of 15% was used. The typical secret key rate was 10 kbps at a 10 dB channel loss. Further, for data processing, we used the Winnow algorithm for error correction [23], and in the privacy amplification process, a shared Toeplitz matrix for key compression was used to obtain secure keys.

In this experiment, the PQC algorithm was integrated into the ARM chip of the QKD device to realize the authentication process. The implementation diagram is shown in Fig. 1.

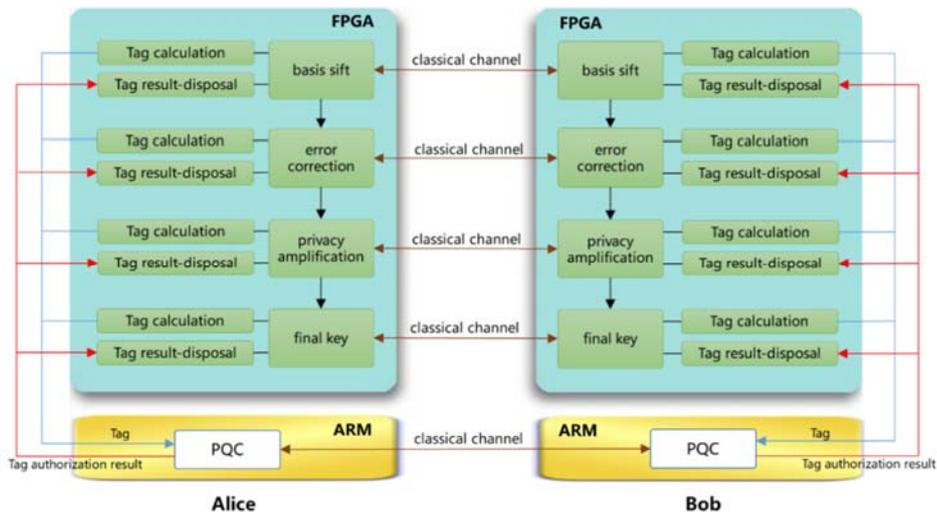

Fig. 1. PQC authentication diagram. The data that need to be authenticated include basis sift data, key after error correction, shared random number in privacy amplification and the final key. The process of data authentication includes: 1) Tag calculation performed through SM3 hash algorithm for data to be authenticated; 2) Tag value sent to PQC module through inter-board communication interface; 3) PQC modules of both communicating parties are authenticated through PQC algorithm; 4) Authentication results return to the module waiting for the authentication result through the inter-board communication interface.

As shown in the figure, Alice and Bob first exchange certificates and random numbers, which are used as nonces to resist replay attacks. Thereafter, the public key of the certificate authority was used to verify the legitimacy of certificates of each other. Next, they used the PQC signature algorithm with its own private key to sign both the message digest and the received random number. Consequently, both parties again used the corresponding public keys to verify the integrity and unforgeability of the signature.

This was implemented on ARM AM3354, the size of public key, private key and signature are 1.3KB, 3.4KB, 2.4KB respectively, and the running time of public and private key generation, signature, and signature verification was less than 10ms. Thus, when the classical channel delay was within 10ms and the bandwidth is not less than 100kbps, the entire process was completed within 100ms.

Further, this experiment was conducted in the Jinan field metropolitan QKD network. The network topology is shown in Fig. 2. There were five optical switching nodes and 14 user nodes in the network.

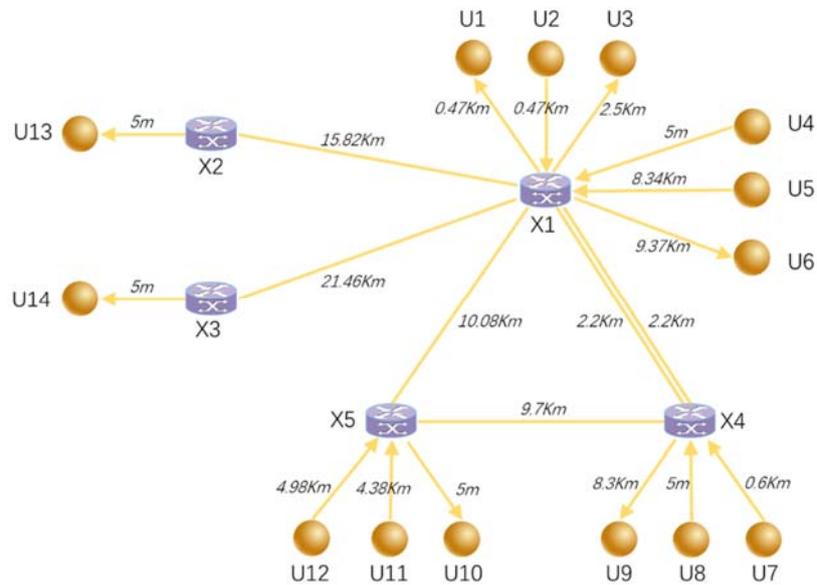

Fig. 2. Jinan field metropolitan QKD network topology, where X1–X5 are optical switching nodes, U1, U3, U6, U9, U10, U13, U14 are receiver QKD nodes, and U2, U4, U5, U7, U8, U11, U12 are sender QKD nodes. Each QKD node is connected to the optical switching node via optical fiber, and the optical switching node is also connected via optical fiber. Based on the actual deployment geographic location of each of the QKD and optical switching nodes, the lengths of optical fibers range from 5 m to 21.46 km.

Before this experiment was performed, the X1, X2, and X3, in the Jinan field metropolitan QKD network, as depicted in Fig. 2, were originally trusted relay nodes. Here, we replaced them with optical switches, reducing the security reliance on the trusted relay in the network, thereby improving the security of the entire QKD network and reducing networking costs. In addition, the network structure and the connection between nodes were simplified due to this change, thereby improving network interoperability.

We measured the signal attenuation of each optical fiber, and then added 1.5dB insertion loss to each optical switching node, thus, we got the signal attenuation of connection between each receiver QKD node and transmitter QKD node, as shown in the Table 1.

Table 1. Signal attenuations of connection between each receiver and transmitter (unit: dB)

| Transmitter / Receiver | U2 | U4 | U5 | U7 | U8 | U11 | U12 |
| --- | --- | --- | --- | --- | --- | --- | --- |
| U1 | 4.1 | 3.1 | 6.6 | 10.37 | 7.46 | 14.47 | 15.27 |
| U3 | 5.8 | 4.8 | 8.3 | 12.07 | 13.57 | 16.17 | 16.97 |
| U6 | 9.5 | 8.5 | 12 | 15.77 | 12.86 | 19.87 | 20.67 |
| U9 | 11.36 | 10.36 | 13.86 | 8.91 | 6 | 15.74 | 16.54 |
| U10 | 11.27 | 10.27 | 13.77 | 11.55 | 8.64 | 4.1 | 4.9 |
| U13 | 10 | 9 | 12.5 | — | — | — | — |
| U14 | 12 | 11 | 14.5 | — | — | — | — |

There are seven receivers and transmitters each in the network, resulting in a total of 7 × 7 = 49 pairs. Among them, 11 pairs cannot generate a key due to high attenuation of the optical fiber (indicated in red). In addition, three groups of optical switches exist in the network, and due to the functional limitation of key management system (KMS), automatic scheduling of more than three sets of optical switches on the same fiber link is not supported. Further, a total of 8 connections from receivers U7, U8, U11, U12 to transmitters U13, U14 could not be paired for key generation. However, the remaining 49 – 11 – 8 = 30 connections successfully generated the keys and thus, were kept running.

We analyzed the running log file of QKD devices, extract the key rate and QBER of each QKD pairing process. We estimate the standard deviation of each connection's key rate data, and eliminate the data with more than three standard deviations, finally, we calculated the average by using statistical average method, then got the key rate of each connection during the experiment period. In the QKD devices of this experiment, when the real-time QBER exceeds the threshold (3.125%), the related key data will be discarded, moreover, if the threshold is exceeded three times in a row, the QKD device will start the optical self-calibration process. Thus, we ignored the QBER that exceed the threshold and calculate the average value of the remaining data by using statistical average method, then got the QBER of each connection in normal pairing process during the experiment period.

The average key rate and QBER of each connection during the 36 days of network operation, are presented in Table 2.

**Table 2. Key rate and QBER of each connections**

| Connection | Route | Length(Km) | Loss(dB) | Pairing process times | Key rate(kbps) | QBER |
|---|---|---|---|---|---|---|
| U2-U1 | U2-X1-U1 | 0.94 | 4.1 | 128 | 16.437 | 1.204% |
| U2-U3 | U2->X1->U3 | 2.97 | 5.8 | 130 | 15.544 | 1.106% |
| U2-U6 | U2->X1->U6 | 9.84 | 9.5 | 115 | 9.003 | 1.006% |
| U2-U9 | U2->X1->X4->U9 | 10.97 | 11.36 | 47 | 3.939 | 1.008% |
| U2-U10 | U2->X1->X5->U10 | 10.55 | 11.27 | 114 | 3.201 | 1.068% |
| U2-U13 | U2->X1->X2->U13 | 16.29 | 10 | 90 | 2.91 | 1.623% |
| U2-U14 | U2->X1->X3->U14 | 21.93 | 12 | 70 | 7.11 | 0.865% |
| U4-U1 | U4-X1-U1 | 0.47 | 3.1 | 161 | 35.277 | 0.627% |
| U4-U3 | U4->X1->U3 | 2.5 | 4.8 | 170 | 29.997 | 0.646% |
| U4-U6 | U4->X1->U6 | 9.37 | 8.5 | 136 | 22.094 | 0.633% |
| U4-U9 | U4->X1->X4->U9 | 10.5 | 10.36 | 81 | 8.334 | 0.564% |
| U4-U10 | U4->X1->X5->U10 | 10.08 | 10.27 | 131 | 8.603 | 0.617% |
| U4-U13 | U4->X1->X2->U13 | 15.82 | 9 | 93 | 8.471 | 0.667% |
| U4-U14 | U4->X1->X3->U14 | 21.46 | 11 | 81 | 15.87 | 0.564% |
| U5-U1 | U5->X1->U1 | 8.81 | 6.6 | 119 | 14.913 | 0.825% |
| U5-U3 | U5->X1->U3 | 10.84 | 8.3 | 155 | 16.647 | 0.601% |
| U5-U6 | U5-X1-U6 | 17.71 | 12 | 123 | 8.672 | 0.581% |
| U5-U10 | U5->X1->X5->U10 | 18.42 | 13.77 | 116 | 2.746 | 0.75% |
| U5-U13 | U5->X1->X2->U13 | 24.16 | 12.5 | 101 | 2.684 | 1.218% |
| U7-U1 | U7->X4->X1->U1 | 3.27 | 10.37 | 115 | 3.928 | 0.853% |
| U7-U3 | U7->X4->X1->U3 | 5.3 | 12.07 | 125 | 4.822 | 0.734% |
| U7-U9 | U7-X4-U9 | 8.9 | 8.91 | 60 | 2.669 | 0.843% |
| U7-U10 | U7->X4->X5->U10 | 10.3 | 11.55 | 112 | 3.538 | 0.699% |
| U8-U1 | U8->X4->X1->U1 | 2.67 | 7.46 | 123 | 6.395 | 0.806% |
| U8-U3 | U8->X5->X1->U3 | 4.7 | 13.57 | 129 | 8.383 | 0.611% |

| Connection | Route | Length(Km) | Loss(dB) | Pairing process times | Key rate(kbps) | QBER |
|---|---|---|---|---|---|---|
| U8-U6 | U8->X4->X1->U6 | 11.57 | 12.86 | 121 | 4.3 | 0.557% |
| U8-U9 | U8->X4->U9 | 8.3 | 6 | 83 | 4.857 | 0.763% |
| U8-U10 | U8-X4-X5-U10 | 9.7 | 8.64 | 118 | 4.619 | 0.807% |
| U11-U10 | U11->X5->U10 | 4.38 | 4.1 | 118 | 14.902 | 0.741% |
| U12-U10 | U12->X5->U10 | 4.98 | 4.9 | 142 | 15.482 | 0.613% |

In theory, the higher the link attenuation, the lower is the key rate; however, when considering actual test data, it was found that certain connections simultaneously have higher link attenuation and key rates; for example, the U8–U3 link attenuation (13.57 dB) is greater than that of U8–U9 (6 dB), and the key rate of U8–U3 (8.383 kbps) is also higher than that of U8–U9 (4.857 kbps), which is due to the difference in the performance of the existing QKD devices in the Jinan field metropolitan QKD network.

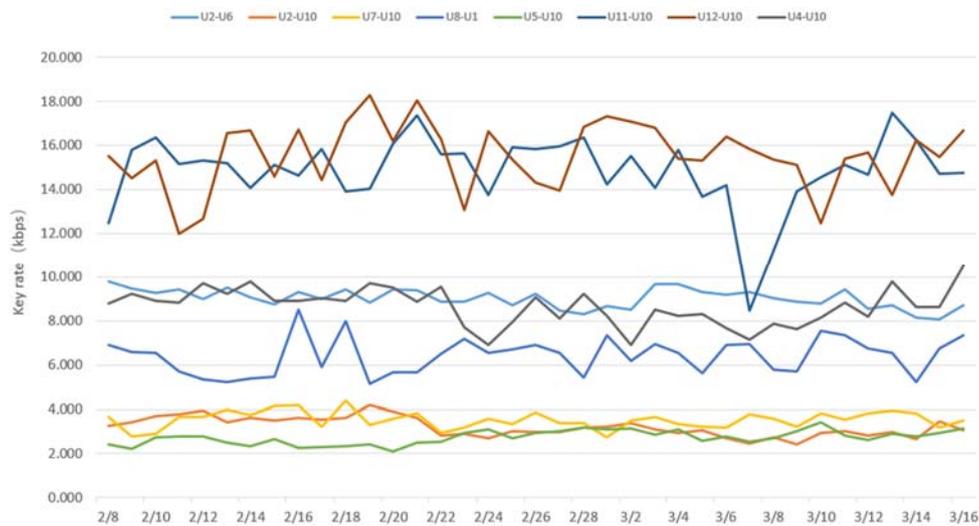

Fig. 3 The change relation between typical connections key rate and time

The experiment was running continuously for more than 36 days, and the entire network maintained a normal operation. Each connection was constantly paired to generate keys, which verified the long-term stability of the PQC algorithm used in the metropolitan QKD network. From 2/8/2021 to 3/16/2021, the average key rate of each connection was counted every day. Considering eight typical connections U2–U6, U2–U10, U7–U10, U8–U1, U5-U10, U11-U10, U12-U10, and U4–U10 as examples, the key rate changes over time are shown in Figure 3.

Furthermore, in order to compare the key rate of using pre-shared key and PQC algorithm authentications in a similar hardware environment and by using same optical fiber, we restored the software of U4-U3 connection to the version of using pre-shared key authentication, kept running for more than 6 hours, and then calculated the average key rate. The corresponding experimental results are presented in Table 3.

Table 3. Key rate under different authentication methods

| Connection | Authentication Method | Key rate(kbps) |
| --- | --- | --- |
| U4-U3 | pre-shared key | 30.441 |
| | PQC algorithm | 29.997 |

The results show that the key rate of using PQC algorithm authentication is similar to that of using pre-shared key authentication. After integrating PQC algorithm, the delay of tag authentication process does not affect QKD.

In the experiment, we designed a business-driven QKD key matching scheduling strategy and verified it using a U4–U3 connection. First, we changed the key queuing strategy on the KMS to queue according to the number of keys, setting it to queue every 30 min. At the initial moment, on the KMS, the number of paired user keys of U4–U3 connection was shown to have reached 32 MB, and thereafter the key consumption rate of U3 and U4 user devices (quantum security router device) was configured to 64 KB every 1 s to quickly consume cache keys in the QKD devices; this process continued till the KMS showed that the pairing keys amount of U4–U3 connection reached 0 bytes. Subsequently, the QKD network switched the pairing relationship to the U4–U3 connection according to the queuing strategy. By analyzing the running log of QKD device, we observed that the key rate of U4-U3 connection during the experiment was 25.951kbps, which was less than the key consumption rate. Meanwhile, we observed that the U4-U3 connection maintained the pairing during the seven QKD key generation periods to generate the key. The experimental results showed that the KMS queues up according to the number of keys, and this confirmed the effectiveness of the strategy of scheduling QKD node pairing for key generation.

## 3. Conclusion

In summary, this study verified the feasibility of integrating the PQC algorithm into a commercial QKD device. In addition, the feasibility and stability of using PQC for authentication in real and complex network topologies were verified in the Jinan field metropolitan QKD network with 14 QKD nodes and 5 optical switching nodes. In the experiment performed for PQC authentication, we replaced the trusted relay in the Jinan field metropolitan QKD network with an optical switch, which simplified the network topology, and thus, reduced the trusted relay node device in the network. Moreover, this change reduced the security dependence of the trusted relay in the network, improved the safety of the metropolitan area QKD, reduced the cost of the device in the construction of the metropolitan QKD network, and improved the interoperability between the QKD nodes.

Further, in metropolitan QKD network applications, compared to pre-shared key authentication, PQC authentication has no effect on the performance of key rate, avoided the process of pre-shared keys between QKD nodes or between QKD nodes and trusted relays, and had distinct advantages in operability, efficiency, security, and device cost. It was found that in the metropolitan QKD network, when the distance between the two parties of QKD exceeded the point-to-point tolerable distance, the trusted relay could not be replaced by an optical switch, and the QKD receiver and transmitter, could not be paired to generate the key. Furthermore, the application of multilevel optical switches was also limited by the performance of the KMS and other devices. We also verified the long-term stability of using PQC authentication in an actual metropolitan QKD network.

Future improvements will target higher performance of PQC algorithm integrated with QKD device [10], higher completeness of the KMS to support multilevel optical switches and higher Integration of QKD device to enable transmit and receive with the same device, further improve the usability and overall performance of using PQC authentication in an actual metropolitan QKD network.

**Funding.**

National Key R&D Program of China (2017YFA0303901); Science and Technology Major Projects in Anhui Province (16030901056, 1703090101); Yunnan Fundamental Research Project (202001BB050028); Key R&D Plan of Shandong Province (2019JZZY010205); Taishan Scholar Program of Shandong Province; Leading talents of Quancheng industry; Key Program of Special Development funds of Zhangjiang National Innovation Demonstration Zone (ZJ2018-ZD-009).